# Linux for Everyone:
# Can Standardization Drive Mainstream Adoption?

*Rohit J Nandha*[*], *Ronak D Patel*[*]

[*]*Department of Computer Science and Engineering, Parul Institute of Engineering and Technology, Vadodara, India*

[1]*210303105280@paruluniversity.ac.in,* [2]*2203031057098@paruluniversity.ac.in*

## Abstract

Linux, despite its technical superiority and flexibility, remains a niche OS in the consumer markets. Because of fragmentation stemming from diverse distributions, it lacks the standardized experience which discourages mainstream adoption. This foundational paper explores whether a balanced approach to standardization can bridge this gap without compromising Linux's core philosophy of freedom and openness. We analyze historical attempts at unification, such as Flatpak, Wayland, and Snap, identifying reasons for their limited success. Using case studies and statistical insights, we understand how fragmentation affects developers, designers, management users, and gaming-users. The paper proposes a standardized yet modular Linux ecosystem ensuring adaptability for new users and flexibility for power users. Rather than giving a technical solution, this paper discusses the feasibility of a unified Linux experience by providing the groundwork for structured standardization, we aim to inspire future research as well for positioning Linux as a viable alternative to Windows and MacOS without sacrificing its open-source nature.

## 1 Introduction

Linux is a powerful, flexible OS used in various domains such as enterprise, cloud computing, embedded systems, etc [1]. Despite having technical superiority, Linux has faced challenges with mainstream adoption in the markets where windows and MacOS are major dominating players [2]. The major problem behind this fragmentation is because of having multiple distributions, package managers, and desktop environments [3]. While this diversity enables innovation, it also results in compatibility challenges, usability issues, and standardization problems, which makes Linux less appealing for normal users [4].

In the past, Several initiatives aimed at standardizing Linux—such as the Linux Standard Base (LSB) [5], Wayland, which was developed primarily to modernize the Linux graphics stack by eliminating the fragmentation inherent in the legacy X11 system, and Flatpak/Snap for universal application packaging [6]—but it struggled to gain response as supposed. These efforts have been hindered by inconsistencies across distributions, lack of cooperation among stakeholders, and resistance from the open-source community [4].

In this paper, we're trying for a balanced approach for Linux standardization, proposing a modular yet standardized Linux that ensures adaptability for new users and flexibility for the power users. Here, we're not offering any technical solution yet instead, this research serves as a level 0 foundational paper, outlining challenges, feasibility, and potentiality for a more cohesive Linux experience [7]. By understanding past failures of various distributions, we aim to initiate discussion and inspire further research on the standardization of Linux without compromising its open-source philosophy.

### 1.1 Background: The Strength and Weakness of Linux

Linux has established itself as a dominant player in the enterprise, cloud, and embedded systems. Powering major web servers, super computers, and IoT devices[1]. In spite of its security, performance and open-source nature, Linux remains a minor player in the desktop OS market, where Windows and MacOS are dominating [2]. The main reason behind this problem is the fragmentation-multiple distros, which makes Linux less user-friendly for the general users [3].

### 1.2 The Fragmentation Problem: A Barrier to Adoption

Windows and MacOS provides a unified user experience and application ecosystem, while Linux is divided into hundreds of distributions, each with different system architectures, package management systems (APT, DNF, Pacman), and desktop environments (GNOME, KDE, Xfce) [4]. This flexibility encourages innovation for sure, but it also results in software incompatibility, inconsistent user experiences, and increased complexity, discouraging mainstream adoption [3].

### 1.3 Past Standardization Efforts and Their Challenges

To solve this fragmentation problem, several initiatives have attempted to standardize Linux at various levels, for example: Linux Standard Base (LSB) aimed to create a common standard for distributions but failed due to low adoption by major distributions and outdated specifications [5]. Wayland was introduced as a replacement for X11, promising better security and efficiency, but slow adoption, compatibility issues, and lack of industry support make it less attractive [4]. Flatpak and Snap were designed as universal package formats to simplify software distribution, but disagreements between major distri-butions such as Ubuntu was favoring Snap while Fedora and other were favoring Flatpak which as a result prevented unified adoption [6].

*1.4  Need for a Balanced Standardization Approach*

Instead of enforcing a one-size-fits-all solution, we're proposing a hybrid approach to Linux standardization—one through this paper such that:

- It maintains core flexibility for advanced users and developers.
- Provides a more standardized experience for mainstream users.
- Encourages collaboration between major distributions rather than competition.
- By analyzing the successes and failures of past standardization attempts, this paper presents a foundational framework (Level 0 research) that serves as a launchpad for future discussions and technical implementations [7].

# 2  Methodology

This chapter outlines the systematic approach used to investigate the research question. Can standardization drive mainstream Linux adoption without compromising its open-source philosophy? The methodology is in accordance with established research guidelines for structuring academic inquiry and balances qualitative analysis with practical constraints.

*2.1  Research Design*

This research is having a qualitative literature review and conceptual analysis to explore how Linux is facing challenges of fragmentation and the potential for standardization to attract general user. The study that we're putting here focuses on synthesizing insights from existing academic papers, industry reports, and case studies. By examining previous attempts such as Linux Standard Base (LSB) and universal packaging systems like Flatpak and Snap, this paper will identify what went wrong with the user adoption. Additionally, the research discusses efforts like Wayland, which aims to modernize Linux's graphical stack by replacing X11 rather than serving as a broader standardization initiative. The study also contrasts Linux's decentralized model with proprietary OS like Windows and MacOS. This comparison highlights the strengths and weaknesses of Linux's open-source philosophy, which addresses power users' needs and general users' problems.

*2.2  Source Selection and Criteria*

Sources were selected based on relevance, credibility, and currency. Key criteria included:

- **Relevance**: Focus on Linux fragmentation, standardization, and user adoption.
- **Credibility**: Peer-reviewed journals (e.g., [8], [1], [7]), industry reports (e.g., [9], [10], [11]), and reputable blogs (e.g., [3], [4], [12]).
- **Currency**: Sources published within the last decade, with exceptions for seminal works (e.g., [13]).

Key references include:

- **Market data**: StatCounter [9] and 6sense [11] for desktop/server market share.
- **Technical analyses**: Linux Foundation [6], Silakov [14], and Jin et al. [15] on standardization challenges.
- **User perspectives**: Musunuru [2], Alam [16], and Reddit discussions [**ref27**] on adoption barriers.

*2.3  Analytical Approach*

This study's analytical approach integrates thematic analysis, comparative analysis, and synthesis to address the research objectives. A comprehensive review of academic literature, industry reports, and community discussions provides the basis for understanding the complex dynamics of the Linux ecosystem.

*2.3.1 :  Thematic Analysis*

The thematic analysis focuses on three stages.

1. **fragmentation challenges:** The problem arises from Linux's decentralized model, and having the existence of multiple desktop environments such as GNOME and KDE leads to inconsistent user experience, which makes it harder to learn for new users [3]. In addition, the presence of various package management systems like APT, DNF, etc, contributes to discordant formats and dependency conflicts, as mentioned in detail by [15] and [12]. Driver incompatibilities also impede usability for non-technical users, particularly those dependent on proprietary OS such as Windows and MacOS [2]. Moreover, community discussion on platforms like Reddit shows that the immense availability of over 300 distributions creates confusion and choice overload, which makes it hard for mainstream adoption [17].

2. **standardization efforts:** Previously to solve this problem, initiatives such as the Linux Standard Base (LSB), tried to unify Linux distribution but encountered in low adoption due to ingrained distribution-specific customization [14], [5]. Contemporary efforts illustrated by the emergence of universal packaging systems like Snap and Flatpak, which aim to mitigate compatibility issues, still have faced resistance from people who value flexibility [12]. Moreover, enterprise-level fragmentation, as indicated by customer-splits such as RHEL versus AlmaLinux, complicates cross-distribution collaboration and standardization [18].

3. **user impacts:** Despite Linux's dominant role in servers, where market analyses indicating a up to 96.3% top web servers [11], it still remains absent in the desktop market with a market share of approximately 3.71% [9]. This disparity is partly attributed to software gaps, including limited support for popular proprietary applications and AAA gaming titles, which put off casual users [2].

*2.3.2 : Comparative Analysis*

The comparative analysis focuses on Linux's decentralized model with a more standardized ecosystem like Windows and MacOS. Proprietary systems like MacOS impose UI/UX standards that simplify usability, whereas Linux's diverse approach can result in fragmented and inconsistent interactions [19]. Moreover, centralized software distribution platforms exemplified by the Microsoft Store streamlined installation processes, while Linux users must be dependent on various package management systems that require significant technical expertise [20]. Enterprises appreciate Linux for its cost savings and robust security measures [1], [2], while individual users are often attracted by the familiarity and integrated experience offered by proprietary operating systems [2].

*2.3.3 : Synthesis of Findings*

Synthesizing these insights, the study proposes a modular standardized framework that seeks to combine flexibility with usability. The framework supports for a **unified core system**, inspired by the LSB, to make sure compatibility for critical components such as the kernel and drivers [14], [5]. This is a **hybrid package management** strategy that integrates universal formats (e.g., Snap and Flatpak) with distribution-specific tools to diminish dependency conflicts [12]. Moreover, the framework suggests the adoption of **standardized UI/UX guidelines** (similar to those proposed in GNOME's Human Interface Guidelines) that provide a consistent user experience while permitting the necessary customization required to power users [21]. At the end, to maintain backward compatibility, the integration of containers and compatibility layers is required, ensuring that the preexisting applications remain functional in an evolving system [22].

## 2.4 Limitations

Even after a thorough analytical approach, several limitations have been acknowledged. First, the study is preponderantly dependent on secondary data sources, including academic literature, industry reports, and community discussions, which might have their own biases and not be fully able to understand the nuanced experience of all users [2], [17]. Second, market data available about Linux's server and desktop share are derived from snapshots in time, which might be overlooking regional variations and rapid shifts in technology [9], [11]. Third, the Linux ecosystem is distinguished by rapid evolution, driven by frequent kernel updates and the emergence of new technologies such as AI workloads, which may provide some conclusions outdates in a short period [22]. At last, the qualitative synthesis limits the ability to quantitatively validate the report, particularly from influential stakeholders who favor maintaining proprietary extensions, a challenge underscored in recent industry analyses [18], [23].

*2.4.1 : Universal Packaging Systems*

Fragmentation in Linux has been a problem for a long period of time in software standardizations. Traditional package managers such as APT (Debian-based), RPM (Red Hat-based), and Pacman (Arch-based) create inconsistencies, which makes software deployment complex across different distributions. To solve this, universal packaging systems—Flatpak, Snap, and AppImage—have emerged as cross-distribution solutions [12].

- Flatpak, developed by Red Hat, aims to deliver application sandboxing and cross-platform compatibility. It amplifies security by isolating applications from the host system and managing dependencies independently. However, its dependence on large runtimes increases disk space usage and startup times, leading to performance concerns [22].

- Snap, designed by Canonical, offers automatic updates and rollback features but has been condemned for centralized control via the Snap Store, limiting its adoption among non-Ubuntu distributions [6].

- AppImage, in contrast, takes a lightweight approach, combining all dependencies within a single executable file, allowing applications to run without installation. While it warrants high portability, it lacks a centralized update mechanism and integration with system package managers, making it less suitable for enterprise environments [20].

- 

Although these universal package formats have significantly improved software distribution in Linux, they also have produced back-and-forth in performance, security, and adoption. Verifiable studies on package management challenges shows that while they make installation easy, they don't fully eliminate compatibility issues with traditional package managers, leading to mixed adoption rates [15]. Moreover, enterprises remain uncertain about adopting these things due to concerns regarding security and existing infrastructure [18].

Overall, universal packaging systems represents a notable step toward reducing fragmentation in Linux software distribution, but their long-term possibility depends on achieving a balance between standardization and flexibility [7].

## 3 Results

### 3.1 Current State of Linux Adoption

*3.1.1 : Market Share Analysis*

Linux adoption across different computing platforms discloses a noticeably heterogeneous landscape. In the server domain,

Linux has over 96% dominance of the top one million web servers and is used in the world's top 500 supercomputers. This is because of having robustness, scalability, cost-effectiveness, and inherent security advantages [9], [11]. By contrast, in the desktop market, Linux has a 1.32% share because of proprietary operating systems like Windows, approximately 71.91%, and MacOS, 15.02% [9]. The limited desktop perforation is largely due to the dearth of pre-installed Linux distribution on PCs, software compatibility issues, and a fragmented ecosys- tem, which complicates the user experience [16].

In the enterprise sector, Linux has broader accepting rate because of its lower cost of ownership, robust security features, and adaptability to cloud-based and containerized environments [1], [24]. Previous studies have noted that organizations often initiate Linux adoption on a limited scale because of its cost effectiveness and ability to replace obsolete systems [8]. Nevertheless, barriers to widespread consumer desktop adoption persist, particularly due to distribution fragmentation and inadequate OEM support [4], [17].

*3.1.2 : User Segment Insights*

Linux adoption is not the same in between different users, with ongoing trends among developers, gamers, enterprise users, and general users.

**Developers:**
Developers favor Linux because of its customizability, extensive development tools, and lively open-source community. As discussed by Barmavat and Karey [1], the flexible environment offered by Linux focuses on innovation and experimentation. In addition, studies on usability have shown that distributions like Ubuntu and Debian are often favoured by developers for both personal and professional use [16], further reinforcing Linux's reputation as a developer-focused platform.

**Gamers:**
Recently, there have been improvements in the gaming environment, such as Proton-Linux's market share, which remains modest among gamers. Das [4] observes that the limited native support for popular game titles and ongoing challenges with driver compatibility are major barriers. This remains constant with broader observations from community discussions that indicate a careful approach among gamers towards adopting Linux as a main gaming platform [17].

**Enterprise Users and General Consumers:**
In the enterprise sector, Linux is widely adopted for server management, cloud computing, and containerized applications because of its lower cost of ownership, robust security, and scalability [18], [24]. However, for general users, the desktop experience is often hindered by a fragmented ecosystem, sheer learning curves, and limited pre-installation on consumer devices [16]. These challenges are additionally highlighted in the research summaries, which show that while enterprise adoption is well built, general-user adoption lags due to usability and compatibility issues [17].

*3.2 Analysis of Fragmentation Issues*

Linux fragmentation is a complicated challenge from both technical and organizational factors that impede the realization of a unified Linux. One primary contributor is the multiplicity and diversity of desktop environments. Different distributions offer distinct graphical UIs (e.g., GNOME, KDE, XFCE), which could be problematic for users and increased learning curves [19].

Another major aspect of fragmentation is the diversity in package management systems. The coexistence of disparate systems like API, RPM, and Pacman leads to dependency conflicts and compatibility issues. Observed studies on package-to-group mechanism highlight how these differences derange seamless software deployment and complicate system maintenance [15], [22].

Moreover, the absence of universal driver support remains a critical concern. With many hardware manufacturers prioritizing Windows and MacOS, Linux users are often forced to rely on community-driven driver development. This situation results in inconsistent performance of hardware and reliability, further fragmenting the user experience [6], [4].

The decentralized development model is a problem at its core. As noted by Ljubuncic, the "sum of egos" within the Linux community leads individual distributions to pursue distinct visions rather than converging on unified standards [3]. However, initiatives such as the Linux Standard Base (LSB) were introduced to promote compatibility over distribution. Still, their limited adoption endured tension between the standardization-focused audience and the flexibility and innovation-focused audience [5].

*3.3 Historical Attempts at Standardization*

*3.3.1 : Linux Standard Base (LSB)*

The Linux Standard Base (LSB) was established as a collaborative effort to unify the major components of different Linux distributions. By establishing common specifications like file system hierarchy, standardized libraries, and system call, the LSB was intended to increase application portability and reduce the fragmentation in the Linux [5].

Despite its hopeful objectives, the LSB has had a limited impact. Many distributions have been hesitant to accept the LSB in full, preferring their distinct configurations and innovations [7]. Additionally, the increasing evolution of the Linux kernel and its associated software often renders parts of the LSB outdated; thereby, it is losing its relevance over a while. This problem between the hope for uniformity and the decentralized, eccentric Linux community has tightened up the standard's global adoption [5], [7].

*3.3.2 : Wayland vs. X11 Transition*

The transition from X11 to Wayland shows a novel effort to standardize the graphical subsystem in Linux by forwarding the longstanding inefficiencies and security vulnerabilities present in the old X11 protocol. Wayland is engineered to streamline communication between the display server and client applications, thereby reducing latency and increasing performance. Despite these theoretical advantages, migration faces challenges.

A primary obstacle is the deep-rooted reliance on X11 within the Linux ecosystem. Over decades, a range of applications and desktop environments have been developed based on X11, by making transition both technically and logistically become complex [6]. Moreover, while Wayland provides enhanced security and efficiency, its adoption has been pannier by compatibility issues with legacy applications and a fragmented landscape among desktop environments [4]. In several cases, developers and users continue to stick with X11 because of its flexibility, which, despite its imperfections, affords substantial customization capabilities essential for certain workflows [7].

Overall, the Wayland versus X11 transition sums up the broader challenges of standardization in Linux. It shows the friction between the modern aspiration for a unified graphical system and the old-is-gold kind of decentralized practice that has long defined the community.

*3.3.3 : Universal Packaging Systems*

Recent efforts to reduce fragmentation in Linux distribution have led to the development of universal packaging systems such as Flatpak, Snap, and AppImage. These solutions focus on providing a unified package format that operates seamlessly throughout diverse Linux distributions, thereby discouraging long-standing issues of dependency conflicts and installation complexities. Whittaker [12] outlines the evolution of Linux package management as a continuous effort to harmonize distribution-specific practices with the need for cross-platform affinity.

Flatpak, Snap, and AppImage each have distinct approaches. Flatpak focuses on robustness and security, Snap integrates features like automated updates and rollback capabilities, while AppImage emphasizes portability by allowing applications to run without a traditional installation process. However, Finethy [22] notes that these systems also introduce challenges, including performance overheads, potential security vulnerabilities, and difficulties in fully integrating with native package management workflows. Moreover, empirical comparisons of Linux package management systems [20] disclose that although universal packaging has reduced fragmentation significantly, they still struggle to achieve the ideal balance between standardization and operational efficiency.

To conclude, while universal packaging systems encouraging unified software distribution in Linux, their mixed reception underline the persistent bargain between achieving broad compatibility and maintaining optimal performance.

*3.3.4 : Enterprise Linux Splits*

In the enterprise market, Linux distributions have increasingly fragmented as organizations look for alternatives to traditional vendor lock-in. Historically, Red Hat Enterprise Linux (RHEL) has been dominating the market because of its overreaching support services and strong security framework. However, variance in licensing and subscription models have stimulate the emergence of community-driven alternatives like AlmaLinux, Rocky Linux, and SUSE. These splits are result of a desire for more flexibility, lower cost of ownership, and enhanced transparency in software management [1], [18].

Experimental studies suggest that organizations often begin their Linux adoption on a limited scale to measure performance and cost benefits. When they were faced with increasing subscription fees or restrictive policies, many chose to transition to alternative distributions, which can provide similar capabilities [8]. Moreover, while open source standardizations have improved to some extent, they have not adjusted the divergent market strategies of computing vendors. This continuous friction foster an environment where multiple enterprise Linux solutions coexist; each comes up with their own needs and priorities [7].

Overall, the emergence of these alternative distributions underscores a critical response to vendor lock-in—providing organizations with viable, cost-effective options that also empower them with greater control over their IT infrastructure.

*3.3.5 : Counterarguments: The Benefits of Fragmentation*

While fragmentation is criticized for complicating Linux, it still provides significant benefits that highlight the open-source philosophy. One main advantage is the promotion of innovation. The decentralized development model, sometimes described as the "sum of egos" within the Linux community, inspires diverse experimentation and quick emergence of specialized solutions [3]. This edge led to a vibrant ecosystem where multiple distributions address niche requirements and focus on continuous improvement.

Fragmentation also warrants extensive customization. Unlike proprietary operating systems, Linux allows organizations/users to modify it however they like for specific performance, security, or usability goals. This level of flexibility empowers organizations/users to build system that meets their specific requirements, which can not be available in a more standardized environment [19]. For instance, some distributions prioritize bleeding-edge features and rapid iteration, while others emphasize long-term stability and reliability.

In addition, the diversity born out from fragmentation shows the fundamental ethos of open-source software: freedom of choice. By giving a range of alternatives from experimental to enterprise-grade solutions, Linux ensures that users are not confined to a cup-of-tea-for-all kind of model. This plurality not only accommodates various technical preferences but also powers the overall ecosystem by preventing vendor lock-in and promoting community-driven governance [2].

To conclude, however fragmentation poses challenges in achieving uniformity, it empowers innovation, customization, and user freedom-qualities that are intrinsic to the success of the Linux [6], [13].

# 4 Conclusion

This chapter focuses on insights obtained from the literature review, thematic and comparative analysis, and experimental results discussed in chapters 1 to 3. The target is to contemplate a standardized yet modular Linux without compromising the open-source ethos and to assess the implications of such an approach for mainstream adoption.

## 4.1 Summary of Key Findings

The research reveals that fragmentation in the Linux ecosystem—manifested in diverse distributions, package management systems, and desktop environments—poses notable challenges for mainstream adoption. Key findings include:

- The intrinsic flexibility of Linux, while encouraging innovation and customization, results in an inconsistent user experience and compatibility issues [1]. This fragmentation not only complicates system maintenance but also makes it hard to learn for non-technical users [21].

- Previous standardization efforts such as LSB and X11 to Wayland transition have achieved little success. While these initiatives have aimed to unify core system components and improve portability, they have often been undermined by ingrained distribution-specific practices [3], [5].

- Universal packaging systems (e.g., Flatpak, Snap, and AppImage) have decreased some level of fragmentation. However, they also introduced swap in performance, security, and integration with native package managers [24], [25].

- Market analyses indicate a distinct contrast in Linux adoption between server environments and desktops. Linux dominates the server market and enterprise applications because of its robustness and cost-effectiveness, yet its desktop market share is very low, largely owing to usability issues and lack of pre-installation on consumer devices [7], [14].

## 4.2 Answering the Research Question

The central research question—*Can standardization drive mainstream Linux adoption without compromising its core philosophy?*—finds a careful affirmative response. Proof suggests that a balanced, modular standardization approach is feasible:

- By establishing a unified core like the LSB for having critical components like kernel, drivers, and core libraries, compatibility issues can be improved without enforcing rigid structure [3], [5].

- A hybrid package management strategy that combined universal formats with distribution-specific tools can curtail dependency conflicts while maintaining the customization benefits [24], [12].

- Standardized UI/UX guidelines, similar to those promoted by GNOME's Human interface guidelines, can provide a consistent user experience for mainstream consumers while allowing flexibility for power users and developers [22].

So, standardization implemented as a flexible framework rather than a strict protocol can solve fragmentation issues and enhance usability without fraying the open-source philosophy [6], [4].

## 4.3 Implications for Linux Adoption

The adoption of a standardized yet modular Linux ecosystem carries significant implications:

- **Mainstream Appeal:** Improved interoperability and a unified user experience are likely to lower barriers for non-technical users. This may accelerate the penetration of Linux into the consumer desktop market, complementing its established dominance in enterprise and server environments [19], [14].

- **Enterprise and Developer Benefits:** A more cohesive ecosystem can enhance system reliability, reduce maintenance overhead, and streamline software deployment. This is especially pertinent for enterprise applications and developer-centric environments where consistency and performance are critical [8], [20].

- **Innovation and Customization:** Importantly, a modular standardization framework preserves the flexibility that has driven innovation within the Linux community. This balance is essential to maintaining the dynamic and competitive nature of Linux development [13], [21].

## 4.4 Recommendations for Future Research

Future research should adopt a multi-pronged approach to address Linux standardization. First, experimental validation of the proposed hybrid framework is needed. Experiments and real-world case studies should quantitatively assess how a unified core with modular customization affects system performance, user satisfaction, and user productivity [15], [21].

Longitudinal studies are also critical. Tracking adoption trends over extended periods will help determine how incremental standardization impacts both the integration of emerging technologies—such as AI workloads and containerized applications—and the preservation of legacy functionalities [25].

Furthermore, detailed economic and security evaluations must be conducted. Future work should develop models having cost-benefit robustness and risk assessments to tackle how standardization affects the total cost of ownership and system resilience in enterprise and cloud environments [8], [24].

Another promising area is the increment of cross-platform integration. Research should focus on the adoption of universal packaging systems like Flatpak, Snap, and AppImage with native package management practices to ensure smooth interoperability and compatibility of legacy applications [12].

Lastly, iterative user experience studies are required for refining standardized UI/UX guidelines. By gripping frameworks like GNOME's Human Interface Guidelines, future research can develop a well-balanced design that meets mainstream users' needs as well as power users' needs without leaving behind the ethos [22].

Collectively, these research roadways not only validate the feasibility of a modular standardization strategy but also provide actionable insights for future Linux, which will be the more robust, user-friendly, and economically competitive operating system.

Future work should focus on the experimental validation of the hybrid standardization framework through the list of experiments and case studies for maintaining system performance, user satisfaction, and productivity of developers [15], [21]. Longitudinal studies are needed to track how Linux is being adopted over time, especially with trending technologies such as AI, Blockchain, etc [25]. In addition, in-depth economic and security analyses should be conducted to focus on cost-benefit trade-offs in enterprise and cloud environments [8], [24]. Research on enhancing cross-platform integration is also to be considered; this will include confirming universal packaging systems with native package management and ensuring legacy application compatibility [12]. Finally, iterative US studies should purify standardized UI/UX guidelines to bridge the gap between technical standardization and user-likable design [22]. These investigations will help to ensure that standardization advances Linux's mainstream usage without compromising its fundamental philosophy of being open-source.

In conclusion, while challenges are there, the proposed approach to standardization offers a hopeful pathway for reconciling Linux's innate flexibility with the need for a more unified user experience. This balance is critical for facing Linux to giants like Windows and MacOS.

## 5 Acknowledgements

We would like to express our gratitude to every author whom we have referenced and to our internal guides for their insights and expertise. Grateful to the entire Department of Computer Science and Engineering, Parul Institute of Engineering and Technology, for providing an efficient time for writing this work. To the people on Stack Overflow, Reddit, and Quora, on which we have understood a lot of things regarding the problem of lack of standardization in Linux. Thank you to the Linux Open-Source community for their inspiration for rooting a philosophy inside us like once-a-Linux-always-a-Linux.

This work is dedicated to those who believe in the open-source contribution and a novel innovation that can be accessible to all people.